\documentclass[seceq]{ptptex}

\usepackage{graphicx}

\def\beq{\begin{equation}}
\def\eeq{\end{equation}}
\def\beqn{\begin{eqnarray}}
\def\eeqn{\end{eqnarray}}
\def\det{{\mathrm Det}\,}
\def\etal{{\sl et al.\/}}
\def\ie{{\sl i.\ e.\/}}

\title{
Taylor expansions in chemical potential%
}

\author{
Rajiv \textsc{Gavai},
Sourendu \textsc{Gupta}\footnote{By whom the talk was delivered}
and Rajarshi \textsc{Roy}%
}

\inst{
Department of Theoretical Physics,%
Tata Institute of Fundamental Research,\\%
Homi Bhabha Road, Mumbai 400005, India.%
}

\abst{
Properties of QCD at finite chemical potential ($\mu$) are extracted
using Taylor series expansions. The continuum limits of lattice results
are presented.  The result of expanding the free energy density,
\ie, the pressure, to 6th order in the expansion is shown. The Taylor
coefficients of the chiral condensate are also shown. Relations between
various Taylor coefficients are demonstrated. All this information is
utilised to remove various lattice artifacts from the determination of
the Wroblewski parameter in strangeness production.}

\begin{document}

\maketitle

\section{Introduction}

Lattice simulations of QCD are possible whenever the determinant of
the Dirac operator, $M$, is positive. The proof of positivity amounts
to constructing an operator $Q$ such that $QMQ^{-1}=M^\dag$. Once
such a $Q$ has been obtained, the result that $\det M$ is real follows
trivially.  For QCD at zero chemical potential ($\mu=0$), $Q=\gamma_5$.
In the limit when isospin symmetry is exact, and an isovector chemical
potential $\mu_3$ is switched on, an isospin flip is the appropriate $Q$,
thus showing that QCD at finite $\mu_3$ is amenable to direct lattice
simulation \cite{wilc}. A chiral model analysis found the phase diagram in
the $T$--$\mu_3$ plane \cite{son}, and lattice simulations \cite{latmu3}
verified these results soon thereafter.  At finite baryon chemical
potential ($\mu$) there is no such operator, and direct simulations
are impossible. CP symmetry nevertheless dictates that the free energy
remains real, and that normal thermodynamics is obtained \cite{sg}.

Many methods have been developed to explore the interesting physics in the
$T$--$\mu$ plane \cite{mpl}.  Here I discuss high-order Taylor expansions
of the pressure and first results of systematic Taylor expansions of
many other quantities. The first results on the Taylor expansion of the
free energy (pressure) were presented in \cite{gott} and that of meson
correlators in \cite{miya}. This report is in two section--- the first
deals with the Taylor expansion of the pressure and the second with
that of the quark condensate and related quantities. Following these
two sections is a brief summary.

\section{The pressure}

In \cite{eos1,eos2} the following series expansion is utilised for the pressure,
\beq
  P(T,\mu) = P(T,0) + \chi_{uu}(T)\mu^2 + \frac1{12}\chi_{uuuu}(T)\mu^4
                    +\frac1{360}\chi_{uuuuuu}(T)\mu^6 + \cdots
\label{taylor}\eeq
where the pressure, $P$ of a system at temperature $T$ is given in terms
of the free energy, $F$, by $P=-F/V$, where $V$ is the volume.  A similar
expansion is used by the Bielefeld-Swansea group \cite{biswa}. CP
symmetry forces the odd terms in the expansion to vanish \cite{sg}. The
neglected terms are meant to be of higher order in $\mu$. The Taylor
coefficients are the generalised susceptibilities, which are best defined
by introducing one chemical potential for each flavour of quarks. This
is allowed since flavour is a conserved quantum number in QCD. Then the
susceptibilities of order $n$ are defined as the partial derivatives
\beq
\chi_{f_1\cdots f_n} = -\frac1V\,\left.\frac{\partial^n F}
    {\partial\mu_{f_1}\cdots\partial\mu_{f_n}}\right|_{\mu_{f_i}=0},
\label{sus}\eeq
where $\{f_1,\cdots f_n\}$ are flavour labels. It is then clear that
two assumptions have been made in writing eq.\ (\ref{taylor}), first
that we consider the isospin symmetric case with 2 flavours of quarks,
and second that only the diagonal susceptibilities ($f_1=f_2=\cdots=f_n$)
need to be retained. Both these restrictions are non-essential, but they
are justified by the respective observations that flavour asymmetry plays
a negligible role in the problem \cite{asym} and that the off-diagonal
susceptibilities are negligibly small \cite{eos1,eos2}.

We note here that the susceptibilities are not merely of formal interest
as expansion coefficients of the pressure, but also form observables
in their own right. They control fluctuations \cite{fluct} and the
strangeness production rate \cite{strange} in heavy-ion collisions.

The Taylor expansion can be cast into a form which makes clear how good
the approximation is, and when it breaks down. Define $\mu_i^*$ to be
that value of $\mu$ at which the $i$-th term is equal to the $i+2$-nd
(for example, $\mu_2^*=\sqrt{12\chi_{uu}/\chi_{uuuu}}$). Then the
expansion in eq.\ (\ref{taylor}) can be written as
\beq
  \Delta P(T,\mu) = \chi_{uu}(T)\mu^2\left[ 1 +
     \left(\frac{\mu}{\mu_2^*}\right)^2 \left[ 1 +
        \left(\frac{\mu}{\mu_4^*}\right)^2 \bigg[ 1 + \cdots\bigg]\right]\right],
\label{approx}\eeq
where $\Delta P(T,\mu)=P(T,\mu)-P(T,0)$.  This is manifestly well-behaved
if we have $\mu_2^*>\mu_4^*>\cdots$, and equally well-behaved series
of approximations arise by neglecting terms of order $i$ and above
when $\mu\ll\mu_i^*$. Thus term by term improvement of the series
is possible. Another advantage is that each susceptibility (and hence
$\mu_i^*$) is computed at zero chemical potential, whereby the continuum
limit can be (and has been) obtained by the usual techniques of lattice
gauge theory \cite{eos1,eos2}.

The other nice point about the Taylor series expansion emerges when one
examines its failure. If the sequence $\mu_i^*$ tends to a limit $\mu_*$,
then clearly the series fails to converge for $\mu=\mu_*$.  Therefore any
finite $\mu_*$ is an estimate of the nearest phase boundary.  Other series
extrapolation methods can also be used--- for example, Pad\'e analysis.
Results using susceptibilities up to 8th order for QCD with 2 flavours of
light dynamical quarks in the continuum limit will be presented elsewhere.

   \begin{figure}
       \centerline{\includegraphics[scale=0.54]{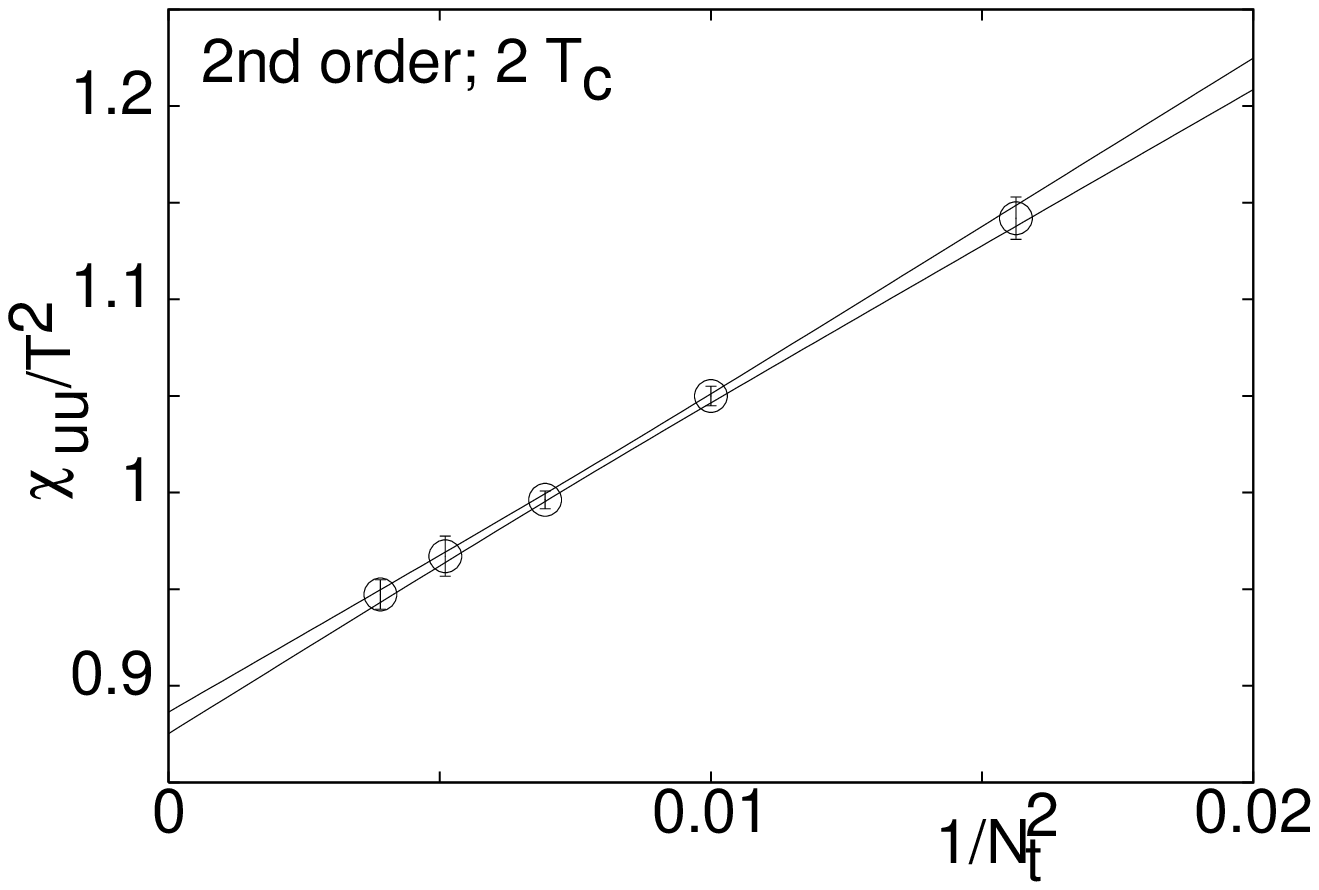}
                   \includegraphics[scale=0.54]{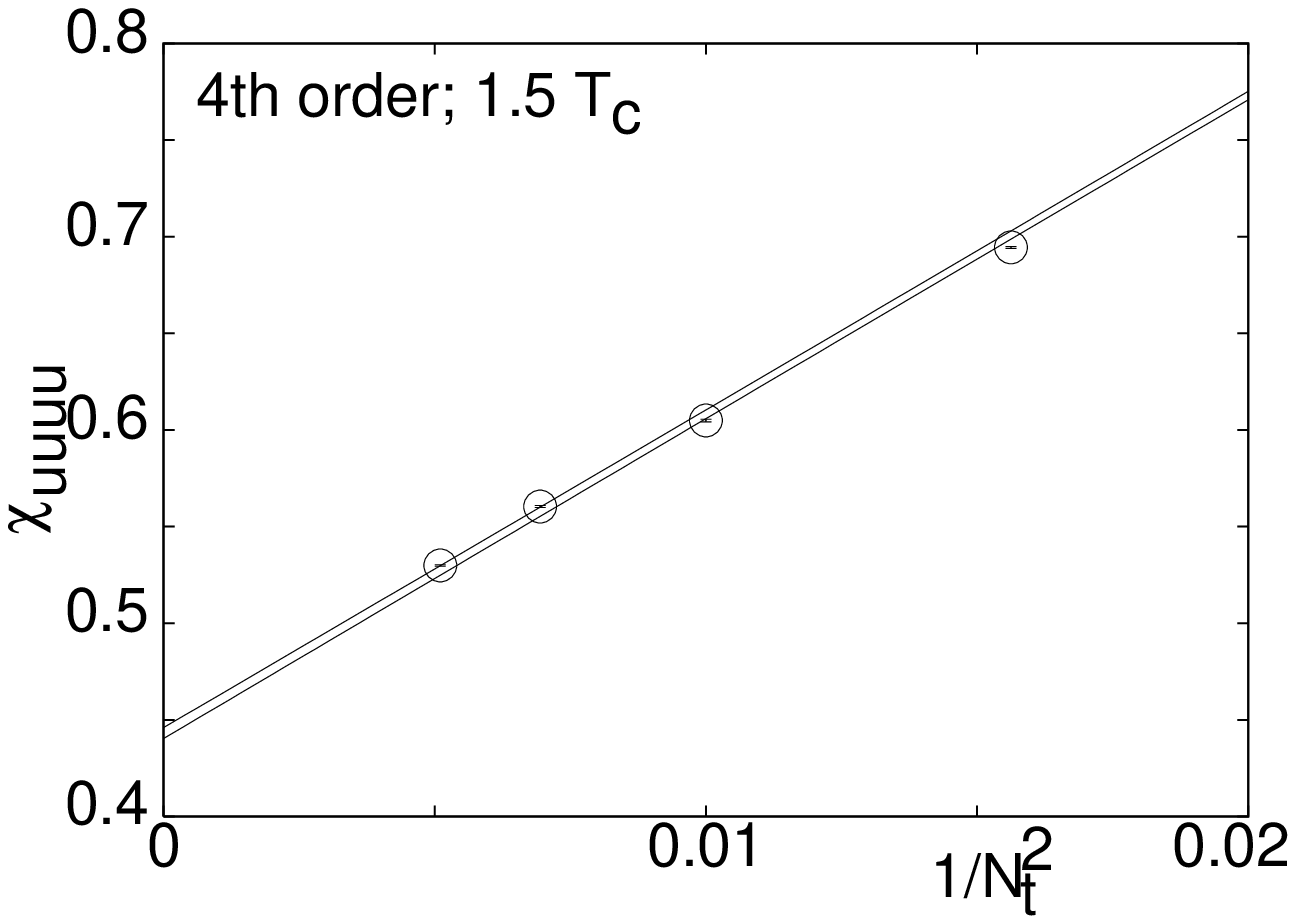}}
       \centerline{\includegraphics[scale=0.54]{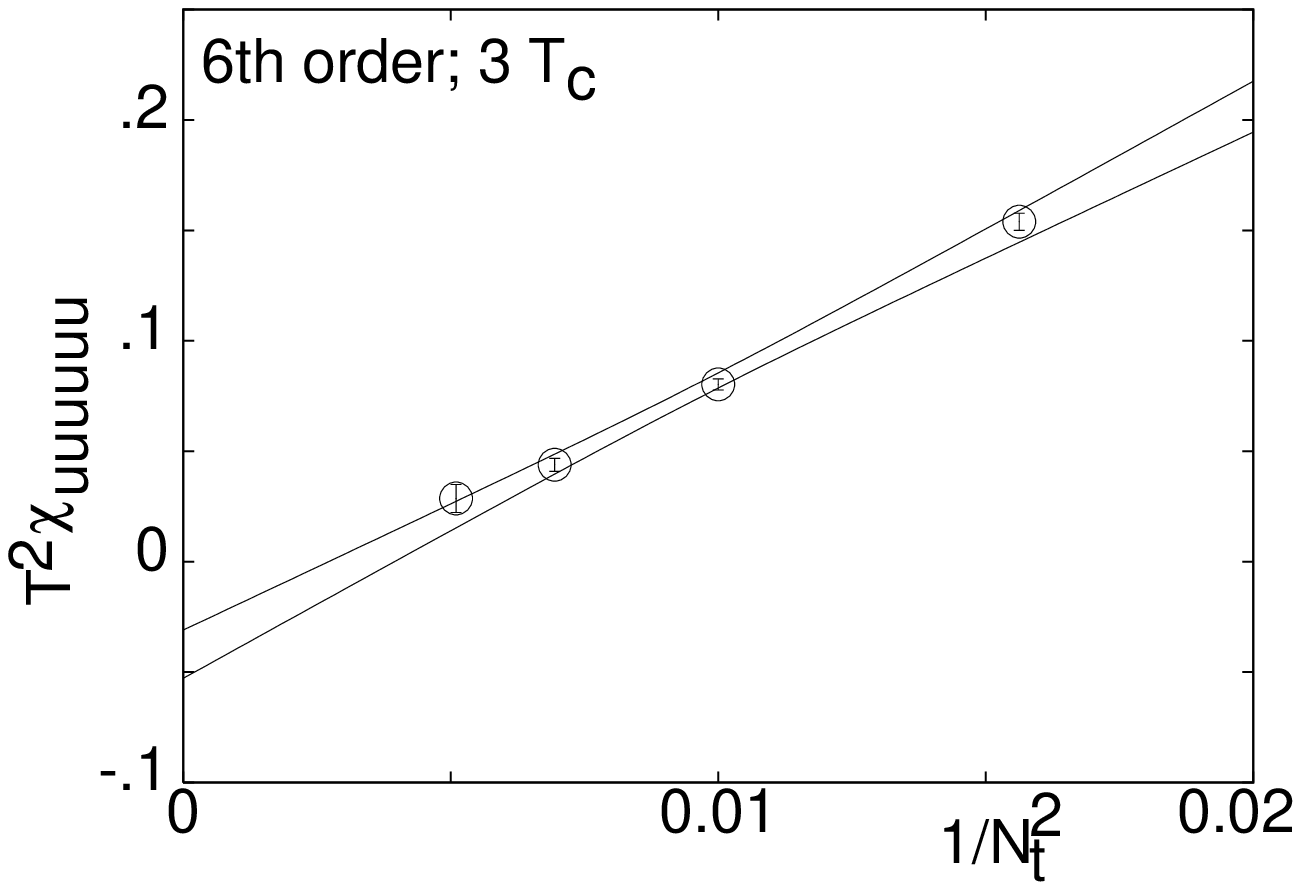}
                   \includegraphics[scale=0.54]{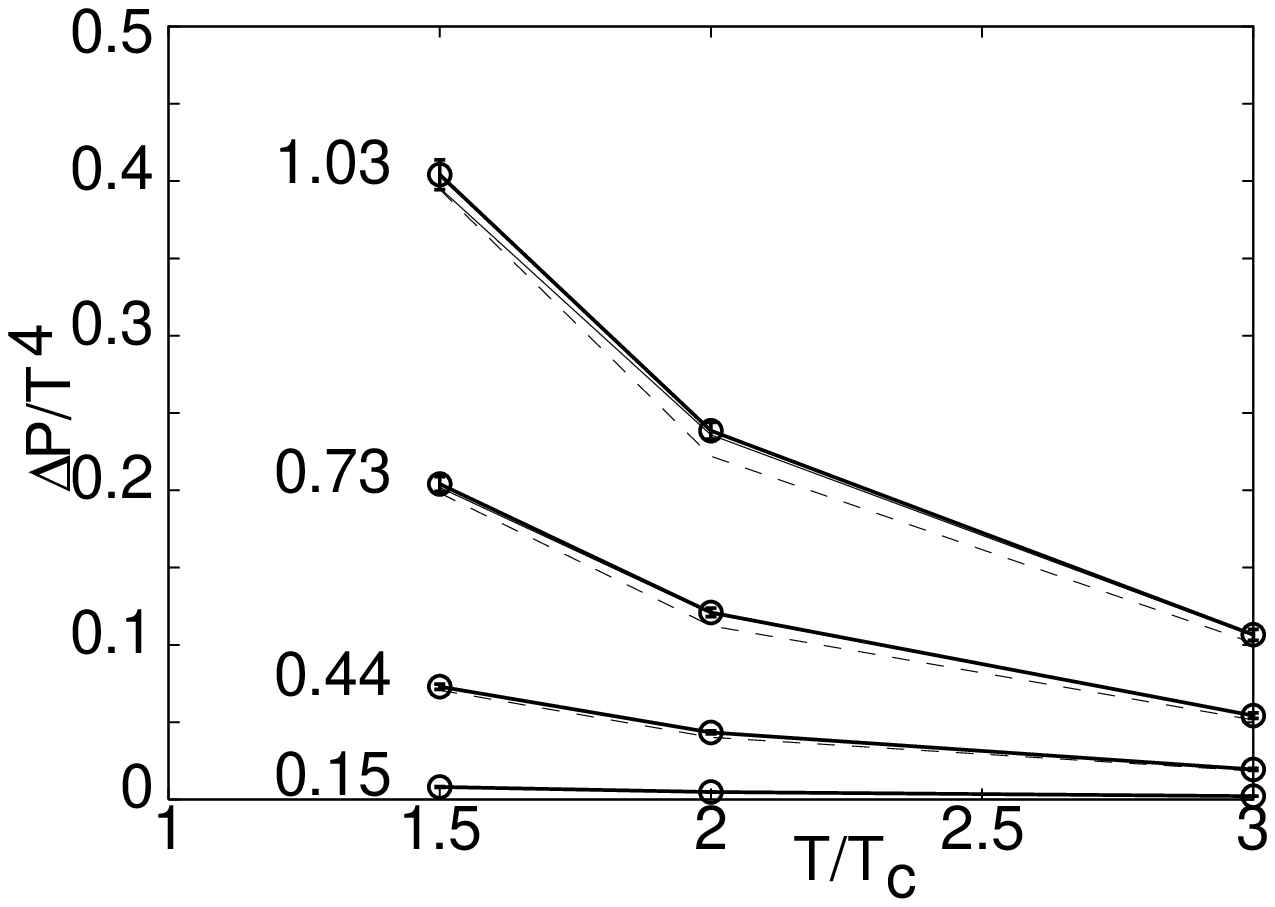}}
       \caption{Continuum limits of the diagonal susceptibilities
          up to 6th order at $T=2T_c$. The bands enclose 66\%
          confidence limits. $\chi_{uuuuuu}$ vanishes at the 99\%
          confidence level. The continuum pressure in high temperature
          QCD for various values of $\mu/T_c$ marked (SPS corresponds
          to $\mu/T_c=0.45$ and RHIC to 0.15).}
   \label{fg.chi}\end{figure}

Like all other lattice methods in current use for this problem, the
Taylor series expansion is limited up to the transition point (or line)
nearest to $\mu=0$. I believe that it has some advantage over other
methods. The primary problem with direct lattice computations at finite
chemical potential is to gain control over phase fluctuations of the
determinant \cite{crompton}. In the reweighting methods this is achieved
by simultaneous movement in both $T$ and $\mu$ \cite{fodor}. If one
could develop a method to sum over CP orbits in configuration space,
then the sign problem would be automatically solved \cite{sg}. In the
Taylor series expansion, one starts from the free energy, where the sum
over CP orbits has already been performed to yield a real quantity.
Thus, an expansion in the single variable $\mu$ completely captures
the physics. The mechanics of the proof involves taking the double
Taylor series expansion of $P$ and checking that it gives no advantage.

In Figure \ref{fg.chi} we display the continuum limits of the diagonal
susceptibilities and the EOS. The latter is compared with the Taylor
series truncated at the second order, and with results obtained through
reweighting on lattices with $N_t=4$ which were extrapolated to the
continuum \cite{fkeos} using the known ratio $\chi_{uu}({\rm cont})/
\chi_{uu}(N_t=4)$ \cite{valence}.  These comparisons show the interesting
fact that in the high temperature phase of the QCD plasma the computation
of the linear susceptibility $\chi_{uu}$ is sufficient to determine
the continuum value of the pressure at $\mu$ of interest to heavy-ion
experiments. This is also evidence that extrapolation in one variable
of $P$ is equivalent to reweighting in two variables.

Not all observable physics in the continuum limit can be extracted by a
simple rescaling by the factor $\chi_{uu}({\rm cont})/\chi_{uu}(N_t=4)$.
At finite chemical potential lattice artifacts are large and connected
to the fact that there is an infinity of equivalent prescriptions for
putting chemical potential on the lattice \cite{gavai}. This results in
an ambiguity in the definition of susceptibilities, and in the critical
end point, which vanishes only in the continuum limit. We have estimated
that this ambiguity may move estimates of the critical end point by an
amount much larger than the present statistical errors \cite{eos1}. Hence,
taking the continuum limit becomes crucial to obtaining this unique
signature of QCD dynamics.

The diagonal linear susceptibilities are in good agreement with
perturbative computations \cite{saclay,drchi}. Consequently,
$\Delta P(T,\mu)$ is also in good agreement with perturbative results
\cite{dreos}.  This is interesting since $P(T,0)$ is badly reproduced
by perturbation theory. Perturbative computations of the fourth order
susceptibility are not as good, and the off-diagonal susceptibilities
fare poorly. It has been suggested that the large $N_f$ behaviour of
perturbation theory can be tested \cite{eos1,largenf} to throw more
light on this.

\section{The quark condensate and meson correlators}

The success of the Taylor expansion of the pressure leads us to believe
that other quantities of interest can also be successfully investigated
at finite chemical potential through Taylor expansions. We use
the quark condensate as an example. This is defined to be
\beq
   C(T,\mu) \equiv \langle \overline\psi\psi\rangle_{T,\mu}
     =\frac1Z\,\frac{\partial Z(T,\mu)}{\partial m}
     =-\frac1{TV}\,\frac{\partial F(T,\mu)}{\partial m},
\label{cond}\eeq
where $Z$ is the partition function, and we have introduced a
non-standard notation, $C$, for the quark condensate which is usually
denoted by $\langle \overline\psi\psi\rangle$. We develop this in a
Taylor series expansion around $\mu=0$,
\beq
   C(T,\mu) = C(T,0) + c_1(T)\mu + \frac12 c_2(T)\mu^2 + \cdots
\label{ctaylor}\eeq
Interesting results follow immediately.

The first derivative at a general value of $\mu$ is
\beq
   c_1(T,\mu) = \frac{\partial C(T,\mu)}{\partial\mu}
              = \frac1T\,\frac{\partial n(T,\mu)}{\partial m},
\label{maxwell}\eeq
where $n(T,\mu)$ is the quark number density, obtained by taking the
first derivative of the pressure with respect to $\mu$. We have related
two seemingly different physical objects
by interchanging orders of derivatives; this is a prototype
of a Maxwell relation in thermodynamics. Further, the Taylor expansion
of the pressure shows that $n(T,\mu) \approx 2\chi_{uu}(T)\mu$, and hence
\beq
   c_1(T,\mu) = \frac{2\mu}T\,
     \frac{\partial\chi_{uu}(T)}{\partial m}+\cdots
\label{maxwellp}\eeq
where the neglected terms start at order $\mu^3$. From this it immediately
follows that $c_1$ in eq.\ (\ref{ctaylor}) vanishes. The outline of a
proof that the Taylor expansion of $C$ in eq.\ (\ref{ctaylor}) is even
is clear from this. Furthermore, it also follows that the limiting value
of $c_1/\mu$ is proportional to the slope of $\chi_{uu}$ with mass.

A Maxwell relation for the second Taylor coefficient $c_2$ follows in a
very similar manner,
\beq
   c_2(T) = \frac1T\,\frac{\partial\chi_{uu}(T)}{\partial m}.
\label{maxwell2}\eeq
This is of great use in lattice computations of strangeness production.
It has been argued that the Wroblewski parameter which can be extracted
from heavy-ion collision experiments on strangeness production can be
written as \cite{strange} $\lambda_s(T) = \chi_{ss}(T)/\chi_{uu}(T)$.
The strange and light quark masses are not direct physical observables,
but are free parameters of QCD which are obtained by fitting the
spectrum of mesons at $T=0$. At present it is hard to perform a lattice
computation at realistic values of the light quark masses. One can then
compute $\lambda_s$ at a value of the light quark mass for which the
computation is feasible and use a Taylor expansion in the quark mass to
extrapolate to the physical quark mass value. The leading term in this
Taylor expansion is then $-Tc_2/\chi_{uu}$.

   \begin{figure}
       \centerline{\includegraphics[scale=1.0]{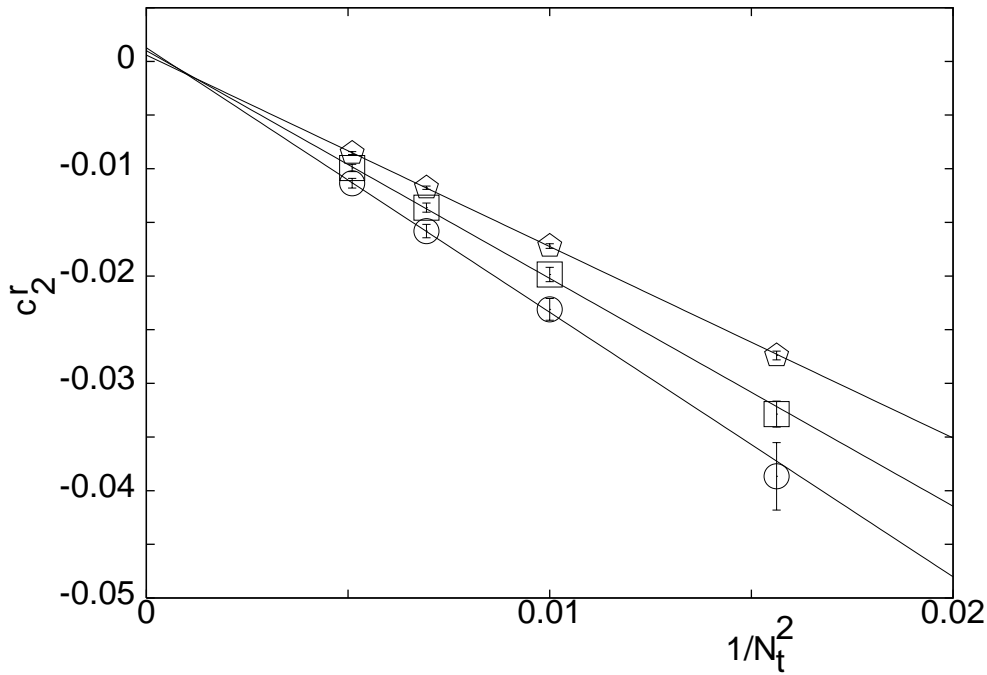}}
       \caption{The renormalised second derivative of the quark
          condensate, $c_2^r$ for $T=1.5Tc$ (circles), $2T_c$ (boxes)
          and $3T_c$ (pentagons). The continuum limit ($1/N_t^2=0$) is
          consistent with zero at the 99\% confidence level at
          all $T$.}
   \label{fg.cond}\end{figure}

One subtlety in these arguments is the effect of renormalisation. Since
$\mu$ appears as a coupling to a conserved charge, it is not renormalised,
and hence renormalisation affects the computations of the susceptibilities
only in so far as they need to be extrapolated to the continuum limit.
The situation is different for the quark condensate and its Taylor
coefficients. The mass is renormalised, by a multiplicative factor for
staggered quarks. This requires a multiplicative renormalisation of every
term in the series in eq.\ (\ref{ctaylor}). This is straightforward. Since
the renormalisation constant depends only on the ultraviolet cutoff,
\ie, the lattice spacing $a$, and not on $T$ and $\mu$, renormalisation
can be accounted for simply by making a Taylor series expansion of
$C(T,\mu)/C(0,0)$. The series remains even, and results for $c_2(T)$ are
shown in Figure \ref{fg.cond}. The renormalised value of $c_2$ is
consistent with zero at the 99\% confidence level.

The Taylor expansion can also be performed in the isovector chemical
potential, $\mu_3$. A comparison of the corresponding Taylor coefficients
reveals that numerically $c_2$ is very nearly equal to the corresponding
coefficient in the series in $\mu_3$ (in contrast, exact equality is
obtained in a random matrix model \cite{klein}). This turns out to be a
fairly general feature, being reproduced also in the Taylor expansions of
the meson correlators. It is due to the observed fact that off-diagonal
susceptibilities like $\chi_{ud}$ are small.

We turn now to correlators of quark bilinears, which are called meson
correlators in extension of their zero-temperature meaning. Taylor
expansions of the correlation lengths have been computed earlier using
an expression in which each correlator is saturated by a single
mass \cite{miya}. We make instead a Taylor expansion of the meson
susceptibilities. In the thermodynamic limit of the low-temperature
phase of QCD where mesons are physical degrees of freedom, these
susceptibilities are proportional to the inverse square of the mass
\cite{old}.

An interesting point is the fact that these meson susceptibilities
can be considered either as a sum over the temporal correlator at all
temporal separation, or as a screening correlator \cite{kogut} summed
over all spatial separation. Above $T_c$ the effective representations of
these two correlators are different \cite{reps}. As a result, the meson
susceptibilities can also be used to extract the dependence on $\mu$
of both masses and screening lengths without any need to use a spectral
function in intermediate steps.

Further relations between coefficients of different Taylor series are
given by chiral Ward identities such as $C(T,\mu)=m\chi_{PS}(T,\mu)$,
where $\chi_{PS}(T,\mu)$ is the pseudo-scalar susceptibility. This
particular identity is quite interesting, as can be seen by
writing out the Taylor expansion of $\chi_{PS}$---
\beq
   \chi_{PS}(T,\mu) = \chi_{PS}(T,0) + \chi_{PS}'(T)\mu 
      + \frac12\chi_{PS}''(T)\mu^2 + \cdots.
\label{pstaylor}\eeq
The chiral Ward identity then allows us to use the arguments already
given for the series expansion of $C(T,\mu)$ to argue that the series
for $\chi_{PS}$ is even.

Furthermore, we obtain the interesting relations
\beq
   \chi_{PS}''(T) = \frac1m c_2(T) = \frac1{mT}\,
     \frac{\partial\chi_{uu}(T)}{\partial m}.
\label{psrelation}\eeq
Recall that for $T>T_c$, the infrared cutoff is the Matsubara
frequency, $\Omega=\pi T$, whenever $m<\Omega$. As a result, the meson
screening masses are given by $2\Omega$ (or its lattice equivalent)
\cite{born,laine}, and is independent of $m$. The susceptibility
$\chi_{uu}$ is equal to the vector meson susceptibility (for staggered
quarks, the identity is with a one-link separated meson). By the previous
argument, this must be independent of $m$ for $m<\Omega$. This is also
seen through explicit computation \cite{pushan}. This explains the
observation that both the quark condensate and the PS susceptibility
are independent of $\mu$ \cite{cohen}, at least to quadratic order in
$\mu$. By the connection established earlier between the variation of
$\lambda_s$ and $c_2$, we therefore conclude that the Wroblewski parameter
is insensitive to the light quark mass, provided that $m<\Omega$.  At the
same time, if the dependence of $\lambda_s$ on the strange quark mass,
$m_s$, is considered, then, since $m_s\approx T_c$, there should be
strong dependence of $\lambda_s$ on $m_s$.

\section{Summary}

We have explored the physics of QCD at finite chemical potential through
systematic Taylor expansions of various quantities. This allows us to
use standard lattice techniques to tackle this problem, and obtain the
continuum limit of every observable. The Taylor coefficients themselves
contain interesting physics, and we have given some examples. There are
numerous relations between them--- we have used a chiral Ward identity
and a Maxwell relation as illustrations. In this report we have shown
how the quark mass dependence of the Wroblewski parameter is related to
physics at finite chemical potential through such relations.


%

\end{document}